\documentclass[graphics,floatfix, nofootinbib,tightenlines,nobibnotes, 18pt, twocolumn]{revtex4-1}

\usepackage{graphicx}
\usepackage[english]{babel}
\usepackage[utf8]{inputenc}
\usepackage{amsmath,amssymb}
\usepackage{amsfonts}
\usepackage{multirow}
\usepackage{pstricks}
\usepackage{float}
\usepackage{subfigure}
\usepackage{color}
\usepackage{epsfig}
\usepackage{slashed}
\usepackage{pst-tools}
\usepackage[colorlinks=true, linkcolor=blue, citecolor=blue, urlcolor=blue]{hyperref}

\begin{document}

\title{Excited doubly heavy baryons production via top-quark decays}

\author{Hong-Hao Ma$^{1,2}$}
\email{mahonghao@pku.edu.cn}
\author{Juan-Juan Niu$^{2,3}$}
\email{niujj@gxnu.edu.cn, corresponding author}
\author{Xu-Chang Zheng$^{4}$}
\email{zhengxc@cqu.edu.cn}

\address{$^{1}$ Center for High Energy Physics, Peking University, Beijing 100871, China}
\address{$^{2}$ Guangxi Key Laboratory of Nuclear Physics and Technology, Guangxi Normal University, Guilin 541004, China}
\address{$^{3}$ Department of Physics, Guangxi Normal University, Guilin 541004, China}
\address{$^{4}$ Department of Physics, Chongqing University, Chongqing 401331, People's Republic of China}

\date{\today}

\begin{abstract}

Within the framework of NRQCD, we calculate the production of the excited doubly heavy baryons $\Xi_{bQ}$ through the semi-inclusive production process $t\rightarrow \langle bQ\rangle[n]\rightarrow \Xi_{bQ}+ \bar {Q} + W^+ $, where $Q= b$ or $c$ quark.
The intermediate diquark state $\langle bQ\rangle[n]$ is in the excited P-wave state, including $[^1P_1]$ and $[^3P_J]$ ($J=$0, 1 or 2) in both color anitriplet state $\mathbf{\mathbf{\overline 3}}$ and color sixtuplet state $\mathbf{6}$, that is, $\langle bc\rangle[^{1}P_{1}]_{\mathbf{\overline 3}/ \mathbf{6}}$, $\langle bc\rangle[^{3}P_{J}]_{\mathbf{\overline 3}/ \mathbf{6}}$, $\langle bb\rangle[^{1}P_{1}]_{\mathbf{\overline 3}}$, and $\langle bb\rangle[^{3}P_{J}]_{\mathbf{6}}$.
We find that the contributions from the P-wave states are about one order lower than the S-wave contributions, and this conclusion is consistent with others.
We also analyze the invariant mass and angle differential distributions, and the theoretical uncertainty from the mass parameters, the transition probability, and the renormalization scale.
Finally, we can expect that about $1.14 \times10^{3-5}$ events of excited $\Xi_{bc}$ and $2.47 \times10^{1-3}$ events of excited $\Xi_{bb}$ can be produced per year at the LHC with $\mathcal{L}$ =$10^{34-36}~\rm{cm}^{-2}~\rm{s}^{-1}$.

\end{abstract}

\maketitle

\section{Introduction}

The 	quark model~\cite{Gell-Mann:1964ewy, Zweig:1964ruk, Zweig:1964jf, DeRujula:1975qlm} theoretically predicts the existence of doubly heavy baryons, whose constituent quarks are two heavy quarks ($b$ or $c$) with one light quark ($u$, $d$, or $s$). 
The study of doubly heavy baryons is an important part of hadron research and also one of the major topics in QCD research for its rich physical information. 
Since the first discovery of the doubly heavy baryon $\Xi_{cc}^{++}$ by the LHCb Collaboration of the Large Hadron Collider (LHC) at the European Organization for Nuclear Research (CERN) in 2017~\cite{LHCb:2017iph}, amounts of works have studied the properties of the doubly heavy baryon. 
At present, the production of doubly heavy baryons have been analyzed by the direct production, such as at the $e^+~e^-$ colliders~\cite{Kiselev:1994pu, Zheng:2015ixa, Jiang:2012jt, Ali:2018ifm}, the hadronic production ~\cite{Berezhnoy:1995fy, Baranov:1995rc, Berezhnoy:1998aa, Chang:2006eu, Chang:2007pp, Chang:2009va, Zhang:2011hi, Wang:2012vj, Chen:2014hqa, Ali:2018xfq}, in deeply inelastic $ep$ scattering~\cite{Sun:2020mvl}, the photoproduction mechanisms~\cite{Baranov:1995rc, Li:2007vy, Chen:2014frw, Bi:2017nzv}, the heavy ion collisions~\cite{Yao:2018zze, Chen:2018koh}, and the indirect production via top quark decays, Higgs decays and $Z$ boson decays~\cite{Niu:2018ycb, Niu:2019xuq, Luo:2022jxq}, etc.

Among the indirect production mechanisms, the top quark decay is intriguing because it is the heaviest known fermion with a mass of 173.0~GeV.
The top quark can participate in all interactions, especially strong interactions, and its decay occurs mainly through weak interactions via the process  $t\to b + W^+$. 
Therefore, the production of doubly heavy baryon through top quark decays is very useful for high precision physics in the electroweak sector.
With the advantage of high luminosity and high collision energy, the LHC or a High Luminosity LHC (HL-LHC) has already become a huge ``top factory'', in which sizable amounts of doubly heavy baryon events through top quark decays can be produced. 
Thus, the top quark decays will be a potentially good platform for studying the indirect production mechanism of the doubly heavy baryons and for searching the $\Xi_{bc}$ and $\Xi_{bb}$ baryons that have remained undetected by experiment so far. 

The nonrelativistic QCD (NRQCD) \cite{Bodwin:1994jh, Petrelli:1997ge} is a very practical theory to deal with the processes involving doubly heavy hadrons. 
According to NRQCD, the doubly heavy baryon can be expanded into a series of Fock states with velocity $v$, where $v$ stands for the typical relative velocity of two heavy quarks in the low-energy interactions. 
The high-energy interactions can be calculated by perturbative QCD. 
To be specific, the production of $\Xi_{bQ}$ via top-quark decays can be factorized into two parts, one is the hard process that produces the intermediate diquark state $\langle bQ \rangle$ ($Q$ stands for $b$ or $c$ quark), which can be calculated perturbatively, and the other is the non-perturbative hadronic process from a $\langle bQ\rangle[n]$ diquark state to a doubly heavy baryon $\Xi_{bQ}$, where $n$ stands for the spin and color quantum number of the intermediate diquark.
The second part can be depicted by a nonperturbative matrix element, which is proportional to the transition probability from the diquark state to the baryon $\Xi_{bQ}$.
The nonperturbative matrix elements can be approximately related to the Schr\"{o}dinger wave function at the origin $|\Psi_{bQ}(0)|$ for S-wave states, and the derivative wave function at the origin $|\Psi_{bQ}^{\prime}(0)|$ for P-wave states.
$|\Psi_{bQ}(0)|$ and $|\Psi_{bQ}^{\prime}(0)|$ can be derived from the experiment or some nonperturbative methods like the potential model~\cite{Kiselev:2000jc, Kiselev:2002iy}, lattice QCD~\cite{Bodwin:1996tg}, or QCD sum rules~\cite{Kiselev:1999sc}.
Because these nonperturbative matrix elements can be treated as an overall parameters, we would concentrated on the essential perturbative part $t\to \langle bQ \rangle[n] +\overline{Q}+ W^+$ for the production of the doubly heavy baryons $\Xi_{bQ}$.
Here, $\Xi_{bQ}$ stands for the doubly heavy baryon $\Xi_{bQq}$, where $Q$ represents the heavy constituent quark $b$ or $c$, and $q$ denotes the light $u$, $d$ or $s$ quark, respectively.  
Thus, the study of the doubly heavy baryons can help us to achieve a deeper understanding of the QCD in both perturbative and nonperturbative regions.
Pioneer investigations on the research of excited P-wave state have been analyzed for the production of mesons~\cite{Chang:2004bh, Yang:2010yg, Liao:2018nab, Liao:2021ifc} and the doubly heavy baryons via $W^+$ boson decays~\cite{Zhang:2022jst}.

In our previous work, the relevant contributions for the production of the doubly heavy baryons from the ordinary S-wave, namely ground state of intermediate diquark states, including $\langle bc\rangle[^{1}S_{0}]_{\mathbf{\bar 3}/ \mathbf{6}}$, $\langle bc\rangle[^{3}S_{1}]_{\mathbf{\bar 3}/ \mathbf{6}}$, $\langle bb\rangle[^{1}S_{0}]_{\mathbf{6}}$, and $\langle bb\rangle[^{3}S_{1}]_{\mathbf{\bar 3}}$, have been discussed. 
There will be about $2.25\times10^{4-6}$ events of $\Xi_{bc}$ and $9.49\times10^{2-4}$ events of $\Xi_{bb}$ produced in one operation year through top quark decays at the LHC or HL-LHC with $\mathcal{L}$ =$10^{34-36}~\rm{cm}^{-2}~\rm{s}^{-1}$.
Therefore, there are enough baryon events can be produced experimentally.
For further precise study and analysis, in the following, we shall study the production of high excited doubly heavy baryons, which come from the excitation of the intermediate diquark state.
Similar with S-wave intermediate diquark, the color state of the excited P-wave can be $\mathbf{\overline 3}$ or $\mathbf{6}$ because of the decomposition of $SU(3)_C$ color group.
The spin state of the excited P-wave can be $[^1P_1]$ or $[^3P_J]$ (J=0, 1, or 2).
However, there are only four spin and color configurations for the P-wave state $\Xi_{bb}$, i.e., $[^1P_1]_{\mathbf{\overline{3}}}$ and $[^3P_J]_{\mathbf{6}}$ due to the symmetry of identical particles in the diquark state.
While there are eight spin and color configurations for the P-wave of $\Xi_{bc}$, $[^1P_1]_{\mathbf{\overline 3}/ \mathbf{6}}$ and $[^3P_J]_{\mathbf{\overline 3}/ \mathbf{6}}$.
All the spin and color configurations of the intemediated diquark states in P-wave, $\langle bc\rangle[^{1}P_{1}]_{\mathbf{\overline 3}/ \mathbf{6}}$, $\langle bc\rangle[^{3}P_{J}]_{\mathbf{\overline 3}/ \mathbf{6}}$, $\langle bb\rangle[^{1}P_{1}]_{\mathbf{\overline{3}}}$, and $\langle bb\rangle[^{3}P_{J}]_{\mathbf{6}}$ would be taken into consideration for a further sound prediction.
A careful study on the excited $\Xi_{bQ}$ production shall be helpful for confirming the contributions from the P-wave states and for further testing of the quark model and NRQCD.

The remaining parts of this paper are organized as follows. In Sec. II, we present the detailed calculation technology for the production of $\Xi_{bQ}$ in P-wave states through top quark decays within the framework of NRQCD. Then the total decay widths, differential distributions and theoretical uncertainty are given in Sec. III. Section IV is reserved for a summary.

\section{Calculation technology}
\label{sec:tech}

\begin{figure}[htb]
  \centering
  \subfigure[]{
    \includegraphics[scale=0.4]{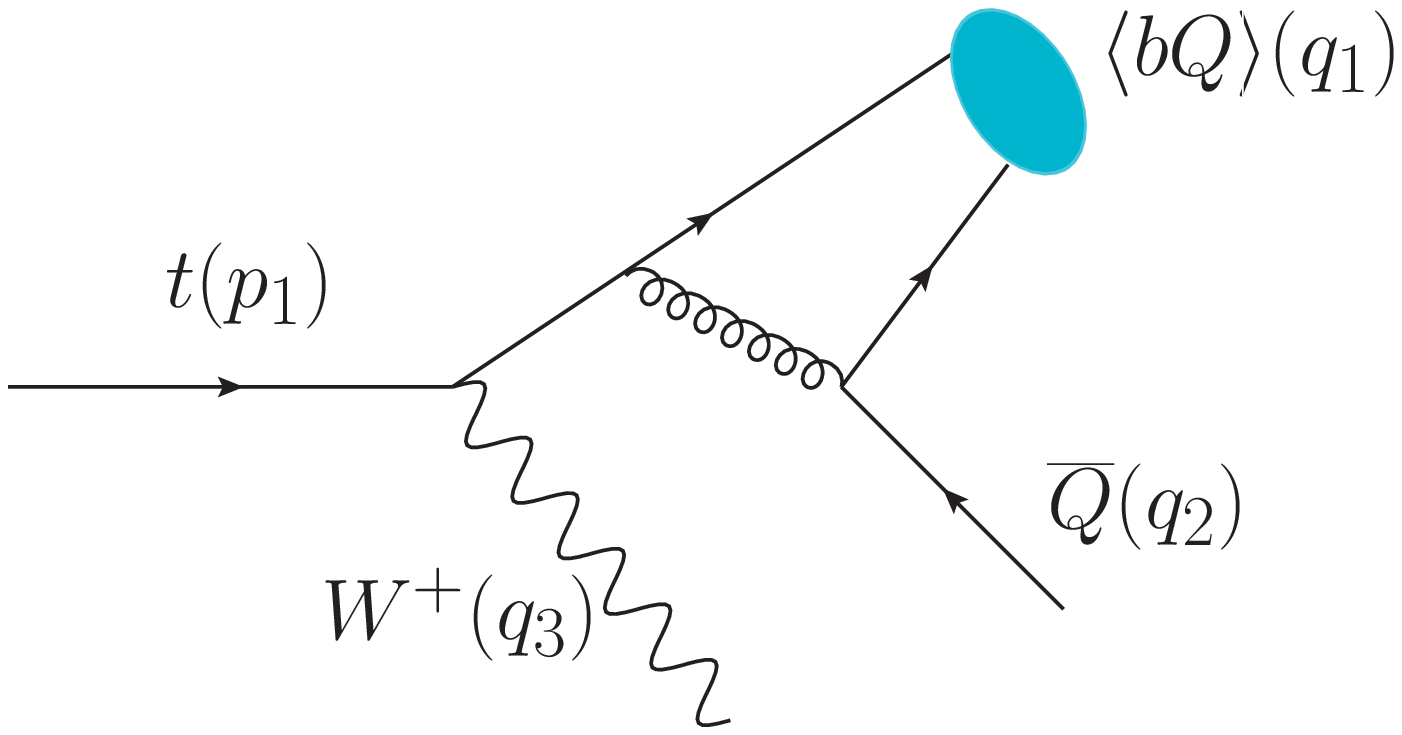}}
  \hspace{0.00in}
  \subfigure[]{
    \includegraphics[scale=0.4]{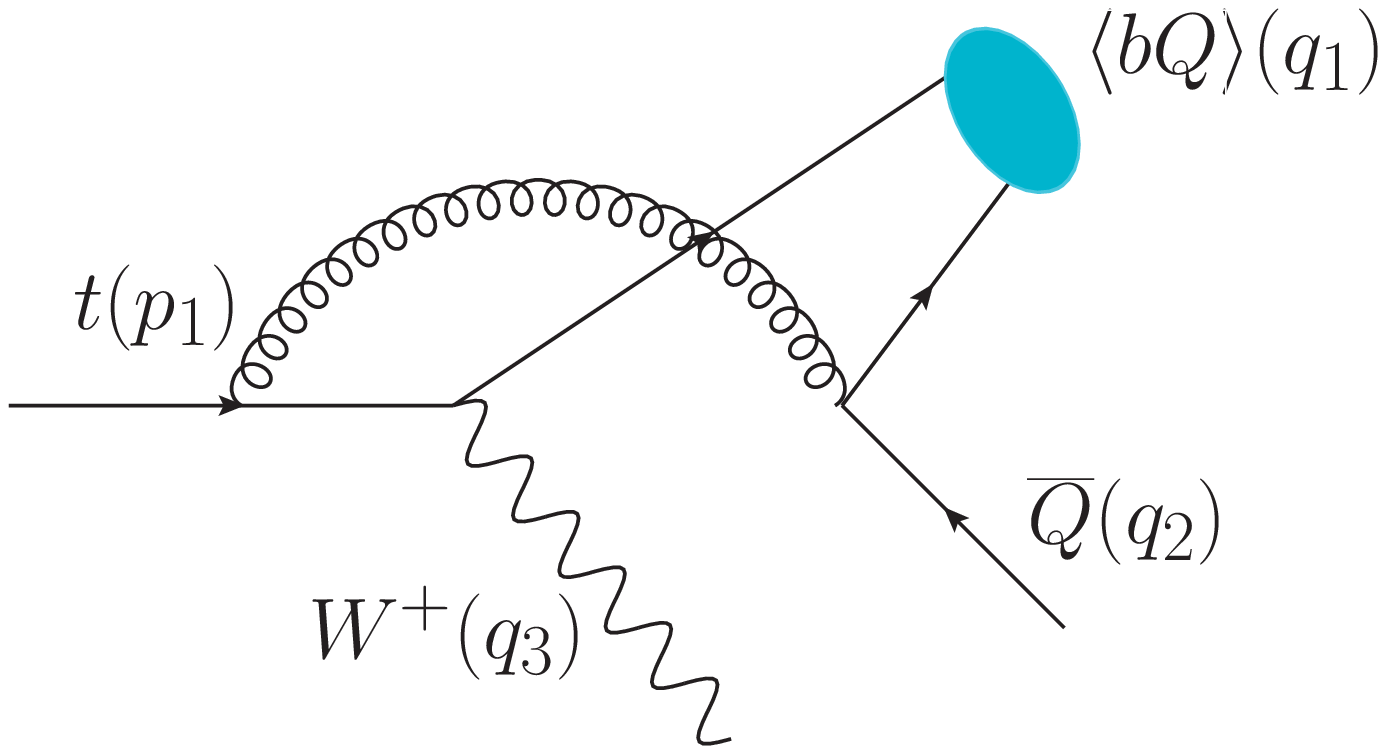}}
  \caption{Typical Feynman diagrams for the process $t(p_1)\rightarrow \Xi_{bQ}(q_1)+ \overline {Q}(q_2) + W^+(q_3) $, where $Q$ denotes the heavy $c$ or $b$ quark for the production of $\Xi_{bc}$ or $\Xi_{bb}$ accordingly.}
  \label{diagram1} 
\end{figure}

Fig.~\ref{diagram1} shows the Feynman diagrams for the process $t(p_1)\rightarrow \Xi_{bQ}(q_1)+ \overline{Q}(q_2) + W^+(q_3)$ .
Within the framework of NRQCD, the decay width can be factorized as~\cite{Bodwin:1994jh, Petrelli:1997ge}:
\begin{eqnarray}\label{factorize}
\Gamma(t(p_1) &\rightarrow& \Xi_{bQ}(q_1)+ \overline {Q}(q_2) + W^+(q_3))  \nonumber \\
&=&\sum_{n} \widetilde{\Gamma}(t \rightarrow \langle bQ \rangle[n] + \overline {Q} + W^+) \langle\mathcal O^{H}[n]\rangle,
\end{eqnarray}
where $[n]$ stands for the spin and color quantum number for the intermediate diquark state, such as  $[^{1}S_{0}]_{\mathbf{\bar 3}/ \mathbf{6}}$, $[^{3}S_{1}]_{\mathbf{\bar 3}/ \mathbf{6}}$, $[^1P_1]_{\mathbf{\bar 3}/ \mathbf{6}}$ or $[^3P_J]_{\mathbf{\bar 3}/ \mathbf{6}}$, and so on. 
All of these different diquark states need to be considered. 
$\langle\mathcal O^{H}[n]\rangle$ represents the nonperturbative matrix element, which is proportional to the transition probability from the diquark state $\langle bQ \rangle[n]$ to the doubly heavy baryon $\Xi_{ bQ}$. 
By relating to the Schr\"{o}dinger wave function at the origin $|\Psi_{bQ}(0)|$ for S-wave states, and the derivative wave function at the origin $|\Psi_{bQ}^{\prime}(0)|$ for P-wave states, one can estimate the value of $\langle\mathcal O^{H}[n]\rangle$.
Specifically, the perturbative short-distance coefficients, that is, the decay width $\widetilde{\Gamma}(t \rightarrow \langle bQ\rangle[n] + \bar {Q} + W^+)$ can be expressed as
\begin{eqnarray}
\widetilde{\Gamma}(t \rightarrow \langle bQ\rangle[n] +\bar {Q} + W^+)= \int \frac{1}{2m_t} \overline{\sum} |\mathcal{M}[n]|^2 d\Phi_3,
\label{width}
\end{eqnarray}
where $\mathcal{M}$ stands for the hard amplitude. 
The meaning of $\overline{\sum}$ is to sum over the colors and spins of all final-state particles ($\langle bQ\rangle$, $\overline {Q}$ and $W^{+}$) and to average over the spin and color of the initial top quark. 
$d\Phi_3$ is the three-particle phase space and it can be represented as
\begin{eqnarray}
d\Phi_3=(2\pi)^4 \delta^4(p_1-\sum_{f=1}^{3} q_i) \prod_{i=1}^{3} \frac{d^3 q_i}{(2\pi)^3 2 q_{i}^{0}}.
\end{eqnarray}
Then we can rewrite Eq.~(\ref{width}) as
\begin{eqnarray}
d \widetilde{\Gamma} = \frac{1}{256 \pi^{3} m_{t}^{3}} \overline{\sum} |\mathcal{M}[n]|^2 ds_{12} ds_{23},
\end{eqnarray}
where $m_t$ represents the mass of top quark, $s_{12}=(q_{1}+q_{2})^{2}$ and $s_{23}=(q_{2}+q_{3})^{2}$ stand for the invariant masses. 
With the help of FeynCalc 9.3~\cite{Shtabovenko:2020gxv} program, we could obtain not only the total decay width but also the corresponding differential distributions.

\subsection{Amplitudes for the diquark production}
As proved in Refs.~\cite{Jiang:2012jt, Zheng:2015ixa}, one can relate the hard amplitude $\mathcal M$ for the production of diquark $\langle bQ\rangle$ with the amplitude of the familiar quarkonium or meson $(b\bar{Q})$ production by applying charge conjugation $C=-i \gamma^2 \gamma^0$ on the heavy $Q$ fermion line.
Remarkably, there is an additional factor $(-1)^{n+1}$ when one use the action of $C$ parity to reverse the heavy $Q$ fermion line and $n$ is the number of vector vertices which equals to one in this process.
By marking the amplitude in the left (right) diagram of Fig. \ref{diagram1} as $\mathcal{A}_1$ ($\mathcal{A}_2$) , one can write the squared hard amplitude as $|\overline{\mathcal{M}}|^{2}=\frac{1}{2 \times 3} \sum |\mathcal{A}_1+\mathcal{A}_2|^{2}$. 
Specifically, the amplitude $\mathcal{A}_1$ ($\mathcal{A}_2$) can be expressed as
\begin{widetext}
\begin{eqnarray}
\mathcal{A}_{1}=i \mathcal{C} \bar{u}_i(q_2, s) \left[\gamma_{\mu} \frac{\Pi_{q_1}[n]}{(q_2+q_{12})^2} \gamma_{\mu} \frac{\slashed{q}_{1}+\slashed{q}_2+m_b}{(q_1+q_2)^2-m_{b}^{2}} \slashed{\varepsilon}(q_3) P_{L}\right] u_j(p_1, s{\prime}), \nonumber \\
\mathcal{A}_{2}=i \mathcal{C} \bar{u}_i(q_2, s) \left[\gamma_{\mu} \frac{\Pi_{q_1}[n]}{(q_2+q_{12})^2} \slashed{\varepsilon}(q_3) P_{L}  \frac{\slashed{q}_{11}+\slashed{q}_3+m_t}{(q_{11}+q_3)^2-m_{t}^{2}} \gamma_{\mu}\right] u_j(p_1, s{\prime}),
\end{eqnarray}
\end{widetext}
in which $\varepsilon(q_3)$ is the polarization vector of $W^{+}$ boson, $P_L= \frac{1-\gamma_{5}}{2}$, the CKM matrix element $|V_{tb}|$ is directly taken to be $\rm{1}$, the overall factor $\mathcal{C}=g g_{s}^{2} \mathcal{C}_{ij,k}$ and the color factor $\mathcal{C}_{ij,k}$ will be explained carefully in Sec. \ref{cf}, $q_{11}$ and $q_{12}$ are the momenta of two heavy quarks, $b$ and $Q$, inside the diquark, respectively, and their forms are:
\begin{eqnarray}
q_{11}=\frac{m_{b}}{M_{bQ}} q_1 +q ~~\rm{and}~~\it{q}_{\rm{12}}=\it{\frac{m_{Q}}{M_{bQ}}  q_1 -q},
\end{eqnarray}
where $q$ is the relative momentum between $b$ and $Q$ inside the diquark, and $M_{bQ}=m_{b}+m_{Q}$ is adopted to ensure the gauge invariance.
The relative momentum $q$ in the diquark state $\langle bQ\rangle[n]$ is small enough to be neglected in the amplitude due to the nonrelativistic approximation.
The projector in the S-wave state can be written as:
\begin{eqnarray}
	\Pi_{q_1}[n]=\frac{1}{2 \sqrt{M_{bQ}}}\varepsilon[n] \left(\slashed{q}_{1}+M_{bQ}\right), \label{piall}
\end{eqnarray}
where $\varepsilon[^1S_0]=\gamma_5$ and $\varepsilon[^3S_1]=\slashed{\varepsilon}$ with $\varepsilon^\beta$ is the polarization vector of diquark in $^3S_1$. 

As to the P-wave amplitudes, the expression can be interconnected with the derivative of the S-wave expression in spin singlet or spin triplet respectively with respect to the relative momentum $q$. 
In order to obtain the derivatives of the projector, $q$ should be preserved and cannot be set directly to zero. Thus the projector can be rewritten as the following forms:
\begin{eqnarray}\label{Psi}
	\Pi_{q_1}[n]=\frac{-\sqrt{M_{bQ}}}{4 m_{Q} m_{b}}\left(\slashed{q}_{12}-m_{Q}\right)\varepsilon[n] \left(\slashed{q}_{11}+m_{b}\right), 
\end{eqnarray}
and it has been proved to be consistent with Eq.(\ref{piall}) when $q=0$~\cite{Zhang:2022jst}.
And then we have to take the derivative of the relative momentum $q$ in the gluon propagator in $\mathcal{A}_{1}$, the gluon propagator and the Fermion propagator in $\mathcal{A}_{2}$. More explicitly, P-wave amplitudes are listed below:
\begin{widetext}
\begin{eqnarray}
\mathcal{A}_{1}[^1P_1]&=& i \mathcal{C} \varepsilon^l_{\alpha}(q_1)\frac{d}{dq_{\alpha}}\left[ \bar{u}_i(q_2, s) \left[\gamma_{\mu} \frac{1}{(q_2+q_{12})^2} \frac{-\sqrt{M_{bQ}}}{4 m_{Q} m_{b}}\left(\slashed{q}_{12}-m_{Q}\right) \gamma^{5}\left(\slashed{q}_{11}+m_{b}\right) \gamma_{\mu} \right.\right.\nonumber\\
 && \cdot \left.\left.\left.\frac{\slashed{q}_{1}+\slashed{q}_2+m_b}{(q_1+q_2)^2-m_{b}^{2}} \slashed{\varepsilon}(q_3) P_{L}\right] u_j(p_1, s{\prime}) \right]\right|_{q=0}, \nonumber\\
\mathcal{A}_{2}[^1P_1]&=& i \mathcal{C} \varepsilon^l_{\alpha}(q_1)\frac{d}{dq_{\alpha}}\left[ \bar{u}_i(q_2, s) \left[\gamma_{\mu} \frac{1}{(q_2+q_{12})^2} \frac{-\sqrt{M_{bQ}}}{4 m_{Q} m_{b}}\left(\slashed{q}_{12}-m_{Q}\right) \gamma^{5}\left(\slashed{q}_{11}+m_{b}\right)\right.\right.\nonumber\\
 && \cdot \left.\left.\left.  \slashed{\varepsilon}(q_3) P_{L} \frac{\slashed{q}_{11}+\slashed{q}_3+m_t}{(q_{11}+q_3)^2-m_{t}^{2}} \gamma_{\mu}\right] u_j(p_1, s{\prime}) \right]\right|_{q=0},
\end{eqnarray}
and
\begin{eqnarray}
\mathcal{A}_{1}[^3P_J]&=& i \mathcal{C}\varepsilon^J_{\alpha\beta}(q_1)\frac{d}{dq_{\alpha}}\left[ \bar{u}_i(q_2, s) \left[\gamma_{\mu} \frac{1}{(q_2+q_{12})^2} \frac{-\sqrt{M_{bQ}}}{4 m_{Q} m_{b}}\left(\slashed{q}_{12}-m_{Q}\right) \gamma^{\beta}\left(\slashed{q}_{11}+m_{b}\right) \gamma_{\mu} \right.\right.\nonumber\\
 && \cdot \left.\left.\left.\frac{\slashed{q}_{1}+\slashed{q}_2+m_b}{(q_1+q_2)^2-m_{b}^{2}} \slashed{\varepsilon}(q_3) P_{L}\right] u_j(p_1, s{\prime}) \right]\right|_{q=0}, \nonumber\\
\mathcal{A}_{2}[^3P_J]&=&i \mathcal{C}     \varepsilon^J_{\alpha\beta}(q_1)\frac{d}{dq_{\alpha}}\left[  \bar{u}_i(q_2, s) \left[\gamma_{\mu} \frac{1}{(q_2+q_{12})^2} \frac{-\sqrt{M_{bQ}}}{4 m_{Q} m_{b}}\left(\slashed{q}_{12}-m_{Q}\right) \gamma^{\beta}\left(\slashed{q}_{11}+m_{b}\right)\right.\right.\nonumber\\
 && \cdot \left.\left.\left.  \slashed{\varepsilon}(q_3) P_{L} \frac{\slashed{q}_{11}+\slashed{q}_3+m_t}{(q_{11}+q_3)^2-m_{t}^{2}} \gamma_{\mu}\right] u_j(p_1, s{\prime}) \right]\right|_{q=0},
\end{eqnarray}
\end{widetext}
where $\varepsilon^l_{\alpha}(q_1)$ ($\varepsilon^J_{\alpha\beta}(q_1)$) is the polarization vector (polarization tensor) of the diquark $\langle bQ\rangle$ in $[^1P_1]$ ($[^3P_J]$ with $J$=0, 1 or 2) state. 
One needs to sum over the polarization vectors or tensors when calculating the squared amplitudes $|\overline{\mathcal{M}}|^{2}$.
For the case of $[^1P_1]$ state, the polarization sum formula is
\begin{eqnarray}
	\sum_{l_{z}} \varepsilon^l_{\alpha} \varepsilon_{\alpha^{\prime}}^{l*}=\Pi_{\alpha \alpha^{\prime}},
\end{eqnarray}
where $l_{z}$ stands for the $[^1P_1]$ states. To obtain the appropriate total angular momentum quantum number $J$  in the case of $[^3P_J]$, we should take the polarization tensor sum formula as
\begin{eqnarray}
	\varepsilon_{\alpha \beta}^{0} \varepsilon_{\alpha^{\prime} \beta^{\prime}}^{0*} &=&\frac{1}{3} \Pi_{\alpha \beta} \Pi_{\alpha^{\prime} \beta^{\prime}}, \label{psum3p0}\\
	\sum_{J_{z}} \varepsilon_{\alpha \beta}^{1} \varepsilon_{\alpha^{\prime} \beta^{\prime}}^{1 *} &=&\frac{1}{2}\left(\Pi_{\alpha \alpha^{\prime}} \Pi_{\beta \beta^{\prime}}-\Pi_{\alpha \beta^{\prime}} \Pi_{\alpha^{\prime} \beta}\right), \label{psum3p1}\\
	\sum_{J_{z}} \varepsilon_{\alpha \beta}^{2} \varepsilon_{\alpha^{\prime} \beta^{\prime}}^{2 *} &=&\frac{1}{2}\left(\Pi_{\alpha \alpha^{\prime}} \Pi_{\beta \beta^{\prime}}+\Pi_{\alpha \beta^{\prime}} \Pi_{\alpha^{\prime} \beta}\right)\nonumber \\
	&-&\frac{1}{3} \Pi_{\alpha \beta} \Pi_{\alpha^{\prime} \beta^{\prime}},\label{psum3p2}
\end{eqnarray}
with the definition 
\begin{eqnarray}
	\Pi_{\alpha \beta}=-g_{\alpha \beta}+\frac{p_{1 \alpha} p_{1 \beta}}{M_{bQ}^{2}}.
\end{eqnarray}

It's worth noting that for the production of $\Xi_{bb}$, the squared amplitudes have to be multiplied by an extra overall factor $(2^2/2!)=2$, where the factor $1/2!$ comes from the identical particles in the $\langle bb\rangle$ diquark state and $2^2$ is the two extra diagrams from the exchange of the two identical quark lines inside the diquark state.

\subsection{Hadronization from diquark $\langle bQ\rangle[n]$ to $\Xi_{bQ}$}\label{hadronization}
 
The hadronization process from the diquark $\langle bQ\rangle[n]$ to the doubly heavy baryon $\Xi_{bQ}$ is nonperturbative.
Within the framework of NRQCD, the nonperturbative matrix element $\langle\mathcal O^{H}[n]\rangle$ can be considered as the transition probability from $\langle bQ\rangle[n]$ to $\Xi_{bQ}$.
For the production of $\Xi_{bQ}$, similar to S-wave intermediate diquark $\langle bQ\rangle$, the color quantum number of the excited P-wave diquark can be also $\mathbf{\bar 3}$ or $\mathbf{6}$ for the decomposition of $SU(3)_{C}$ color group $\mathbf{3}\bigotimes\mathbf{3}=\mathbf{\bar{3}} \bigoplus \mathbf{6}$.
Here $h_{\mathbf{\bar{3}}}$ and $h_\mathbf{6}$ are used to describe the transition probabilities of the diquark $\langle bQ\rangle$ in color antitriplet and color sextuplet, respectively.
Assuming that the potential of the diquark in color $\mathbf{\bar 3}$ is hydrogen-like, $h_{\mathbf{\bar{3}}}$ can be approximatively related to the Schr\"{o}dinger wave function at the origin $|\Psi_{bQ}(0)|$ for S-wave states, and the first derivative wave function at the origin $|\Psi_{bQ}^{\prime}(0)|$ for P-wave states.
And then the first derivative wave function at the origin can naturally connect with the derivative radial wave function at the origin,
\begin{eqnarray}
|\Psi'_{bQ}(0)|^2=\frac{3}{4\pi}|R'_{bQ}(0)|^2.
\end{eqnarray}

While for the transition probability $h_{\mathbf{6}}$, there are two different points of view in the early literatures. One is that due to the ``one-gluon-exchange" interaction inside the diquark, the “binding force”of diquark in color $\mathbf{\overline{3}}$ state is stronger than that in $\mathbf{6}$ state, resulting that $h_{\mathbf{6}}$ shall at least be suppressed by $v^2$ compared to $h_{\mathbf{\bar{3}}}$, or too small to be ignored~\cite{Zheng:2015ixa, Ma:2003zk}.
Here $v$ is the relative velocity between the two heavy quarks inside $\Xi_{bQ}$ in the rest frame of the baryon.
While others indicate that these two parameters $h_{\mathbf{6}}$ and $h_{\mathbf{\bar{3}}}$ are of the same order of $v$~\cite{Ma:2003zk,Chang:2006eu} following the naive NRQCD power counting rule.
Thus, for convenience and to reduce the selection of parameters, we assume $h_{\mathbf{6}}\simeq h_{\mathbf{\overline{3}}}$ in the numerical calculation.

\subsection{The color factor}\label{cf}
For the production of P-wave diquark state $\langle bb\rangle$, there are four spin and color configurations, such as $[^1P_1]_{\mathbf{\bar{3}}}$ and $[^3P_J]_{\mathbf{6}}$ with $J$=0, 1, 2 due to the symmetry of identical particles in the diquark state. 
While there are eight spin and color configurations for $\Xi_{bc}$, $[^1P_1]_{\mathbf{\bar 3}/ \mathbf{6}}$ and $[^3P_J]_{\mathbf{\bar 3}/ \mathbf{6}}$, all these spin and color configurations of the intemediate P-wave diquark states would be taken into consideration for a further sound prediction.
According to the feynman rule, the color factor $\mathcal{C}_{ij,k}$ in Fig. \ref{diagram1} can be calculated by
\begin{eqnarray}
\mathcal{C}_{ij,k}=\mathcal{N} \times \sum_{a,m,n} (T^a)_{mi} (T^a)_{nj} \times G_{mnk}.
\end{eqnarray}
Here $\mathcal{N}=\sqrt{1/2}$ is the normalization constant; $i, j, m, n= 1, 2, 3$ are the color indices of $\overline{Q}(q_2)$, $t(p_1)$, and the two constituent quarks $Q$ and $b$ inside the diquark $\langle bQ\rangle(q_1)$ respectively; $a=1, \ldots, 8$ and $k$ denote as the color index of the gluon and the diquark $\langle bQ\rangle$; When the diquark $\langle bQ\rangle$ is in color $\mathbf{\bar 3}$ or $\mathbf{6}$ state, the function $G_{mnk}$ is equal to the antisymmetric function $\varepsilon_{mnk}$ or  the symmetric function $f_{mnk}$ , accordingly. And they satisfy
\begin{eqnarray}
\varepsilon_{mnk} \varepsilon_{m^{\prime}n^{\prime}k}=\delta_{mm^{\prime}}\delta_{nn^{\prime}}-\delta_{mn^{\prime}}\delta_{nm^{\prime}},
\end{eqnarray}
\begin{eqnarray}
f_{mnk} f_{m^{\prime}n^{\prime}k}=\delta_{mm^{\prime}}\delta_{nn^{\prime}}+\delta_{mn^{\prime}}\delta_{nm^{\prime}}.
\end{eqnarray}
Finally, for the production of diquark in color $\mathbf{\bar 3}$ and $\mathbf{6}$, we obtain the color factors $\mathcal{C}^{2}_{ij,k}$ are equal to $\frac{4}{3}$ and $\frac{2}{3}$, respectively.

\section{Numerical results}
In the numerical calculation, the input parameters are adopted as \cite{Baranov:1995rc, Niu:2018ycb}
\begin{eqnarray}
&&m_c=1.8~\rm{GeV},~~~~~~~\it{m_b}=\rm 5.1~{GeV},\nonumber\\
&&\rm{M}_{\Xi_{\it{bc}}}=6.9~\rm{GeV},~~~~~M_{{\Xi_{\it{bb}}}}=\rm 10.2~{GeV},\nonumber\\
&&|R^{\prime}_{bc}(0)|= 0.200~\rm{GeV^{\frac{5}{2}}},~|\it{R^{\prime}_{bb}}\rm(0)|= 0.479~\rm{GeV^{\frac{5}{2}}},\nonumber\\
&&m_t=173.0~\rm{GeV},~~~~\it{m_W}=\rm 80.385~{GeV}, \nonumber\\
&&\alpha_s(2m_{b})=0.178 , ~~~~\alpha_s(2m_{c})=0.239,\nonumber\\
&&G_{F}=1.1663787 \times 10^{-5}~\rm{GeV^{-2}}, \nonumber\\
&&\it{g}=\rm 2\sqrt{2}\it m_W \sqrt{G_F/\rm\sqrt{2}}, 
\end{eqnarray}
where the masses of constituent quark $m_c$ and $m_b$ are the same as that in Ref.~\cite{Baranov:1995rc}, they are used to build the mass of corresponding baryon $\rm{M}_{\Xi_{\it{bc}}} \simeq \rm{M}_{\it{bQ}}$=$m_{b}+m_{Q}$ ($Q$ = $b$ or $c$). $|R^{\prime}_{bc}(0)|$ and $|R^{\prime}_{bb}(0)|$ are the derivative radial wave functions at the origin, and its values that we use in our calculation are evaluated by the $K^{2}O$ potential motivated by QCD with a three-loop function~\cite{Kiselev:2002iy}. The remaining parameters are the same as that in our previous work, ref.~\cite{Niu:2018ycb}, in order to make a comprehensive comparison and analysis with the contribution of the intermediate S-wave diquark state. The renormalization scale $\mu_r$ for the production of $\Xi_{bc}$ ($\Xi_{bb}$) is set to be $2 m_c$ ($2 m_b$), and the same as the factorization scale. Finally the number of produced $\Xi_{bQ}$ events per year could be roughly estimated by 
\begin{eqnarray}
N_{\Xi_{bQ}[n]}=N_{t} \rm{Br}_{\it bQ}\it [n], \label{Nt}
\end{eqnarray}
where $N_{t}$ represents the total events of top quark produced per year, and there are about $10^{8-10} ~t\bar{t}$ pair~\cite{Kidonakis:2004hr, Kuhn:2013zoa} produced in one operation year at the LHC running with a high luminosity $\mathcal{L}$ =$10^{34-36}~\rm cm^{-2} s^{-1}$. $\rm{Br}_{\it bQ}[n]$ is the branching ratio for the production of $\Xi_{bQ}$, which is defined as 
\begin{eqnarray}
\rm{Br}_{\it bQ}[n]=\frac{\rm{\Gamma}_{\it t\rightarrow \rm{\Xi}_{\it bQ}[n]+ \bar {Q}+W^{+}}}{\rm{\Gamma}_{\it t}}.
\end{eqnarray}

in which [n] is the spin and color quantum number of the intermediate diquark state, $\Gamma_{t}$ stands for the total decay width of the top quark.
The decay width of top quark through the maximum decay channel $t \rightarrow bW^{+}$ is 1.49~GeV, which can be considered as its total decay width.

\subsection{The production of $\Xi_{bc}$ baryon}\label{secbc}
The total decay width of $\Xi_{bc}$ baryon via top quark decays are obtained and listed in Table~\ref{bcwidth}. 
The results contain all considered P-wave intermediate diquark state, and the contribution of the intermediate S-wave diquark states are also listed in Table~\ref{bcwidth} for a comprehensive comparison and analysis. 
S-wave (P-wave) in Table~\ref{bcwidth} stands for the sum of states $[^1S_0]_{\mathbf{\bar{3}}/\mathbf{6}}$ and $[^3S_1]_{\mathbf{\bar{3}}/\mathbf{6}}$( $[^1P_1]_{\mathbf{\bar{3}}/\mathbf{6}}$ and $[^3P_J]_{\mathbf{\bar{3}}/\mathbf{6}}$ with $J$= 0, 1, 2).
\begin{table}[htb]
\caption{Decay widths (KeV) and $\rm{Br}_{\it bc}[n]$ for the production of P-wave and S-wave $\Xi_{bc}$ via top quark decays. States represent the spin and color states for the intermediate diquark.}
\begin{tabular}{|c||c||c|}
\hline
~~State~~&~Decay width~&~$\rm{Br}_{\it bc}[n]$~ \\
\hline\hline
$[^1S_0]_{\overline{\mathbf{3}}}$     & 96.2 & $6.46 \times 10^{-5}$   \\
\hline
$[^1S_0]_{\mathbf{6}}$     & 48.1 & $3.23 \times 10^{-5}$  \\
\hline
$[^3S_1]_{\overline{\mathbf{3}}}$     & 127.6 & $8.56 \times 10^{-5}$   \\
\hline
$[^3S_1]_{\mathbf{6}}$     & 63.8 & $4.28 \times 10^{-5}$   \\
\hline
$[^1P_1]_{\overline{\mathbf{3}}}$     & 2.69 & $1.81\times10^{-6}$ \\
\hline
$[^1P_1]_{\mathbf{6}}$     & 1.34 & $9.03\times10^{-7}$ \\
\hline
$[^3P_0]_{\overline{\mathbf{3}}}$     & 1.82 & $1.22\times10^{-6}$  \\
\hline
$[^3P_0]_{\mathbf{6}}$     & 0.91 & $6.10\times10^{-7}$  \\
\hline
$[^3P_1]_{\overline{\mathbf{3}}}$     & 3.46 & $2.32\times10^{-6}$ \\
\hline
$[^3P_1]_{\mathbf{6}}$     & 1.73 & $1.16\times10^{-6}$ \\
\hline
$[^3P_2]_{\overline{\mathbf{3}}}$     & 3.34 & $2.24\times10^{-6}$  \\
\hline
$[^3P_2]_{\mathbf{6}}$     & 1.67 & $1.12\times10^{-6}$  \\
\hline
S-wave    & 335.7 & $2.25 \times10^{-4}$ \\
\hline
P-wave     & 16.96 & $1.14 \times10^{-5}$ \\
\hline
Total    & 352.66 & $2.36 \times10^{-4}$ \\
\hline
\end{tabular}
\label{bcwidth}
\end{table}
The results indicate that
\begin{itemize}
  \item All of the eight spin and color states for the production of excited $\Xi_{bc}$ baryon have been considered in our numerical calculation and the ratio of the decay widths in these eight excited states is $[^{3} P _1]_{\mathbf{\bar {3}}}:[^{3} P _1]_{\mathbf{6}}:[^{3} P _0]_{\mathbf{\bar {3}}}:[^{3} P _0]_{\mathbf{6}}:[^{3} P _2]_{\mathbf{\bar {3}}}:[^{3} P _2]_{\mathbf{6}}:[^{1} P _1]_{\mathbf{\bar {3}}}:[^{1} P _1]_{\mathbf{6}}=1:0.50:0.53:0.26:0.97:0.48:0.78:0.39$. That is to say, for the production of P-wave $\Xi_{bc}$, the proportions of these eight spin and color states are 24\%, 12\%, 13\%, 6\%, 23\%, 12\%, 7\%, and 3\%, accordingly.

  \item Obviously, the contribution from the spin and color state $[^{3} P _1]_{\mathbf{\bar {3}}}$ domains. The total decay widths with the intermediate diquark in color antitriplet state shall be about twice of that in color sextuplet due to the color factor.
  
  \item The total contribution from intermediate P-wave diquark state is about 5.05\% of that from S-wave.

  \item When all the considered intermediate S-wave and P-wave diquark states are summed, there are about $2.36 \times 10^{4-6}$ $\Xi_{bc}$ events/yr produced via top quark decays at the LHC, about $1.14 \times 10^{3-5}$ events/yr of which came from the P-wave states. The events per year in $[^{1} P _1]_{\mathbf{\bar {3}}}$, $[^{1} P _1]_{\mathbf{6}}$, $[^{3} P _0]_{\mathbf{\bar {3}}}$, $[^{3} P _0]_{\mathbf{6}}$, $[^{3} P _1]_{\mathbf{\bar {3}}}$, $[^{3} P _1]_{\mathbf{6}}$, $[^{3} P _2]_{\mathbf{\bar {3}}}$ and $[^{3} P _2]_{\mathbf{6}}$ states are about $1.81 \times 10^{2-4}$, $9.03 \times 10^{1-3}$, $1.22 \times 10^{2-4}$, $6.10 \times 10^{1-3}$, $2.32 \times 10^{2-4}$, $1.16 \times 10^{2-4}$, $2.24 \times 10^{2-4}$ and $1.12 \times 10^{2-4}$, accordingly.
\end{itemize}

\begin{figure}[htb]
  \centering
  \subfigure[]{
    \includegraphics[width=0.47\textwidth]{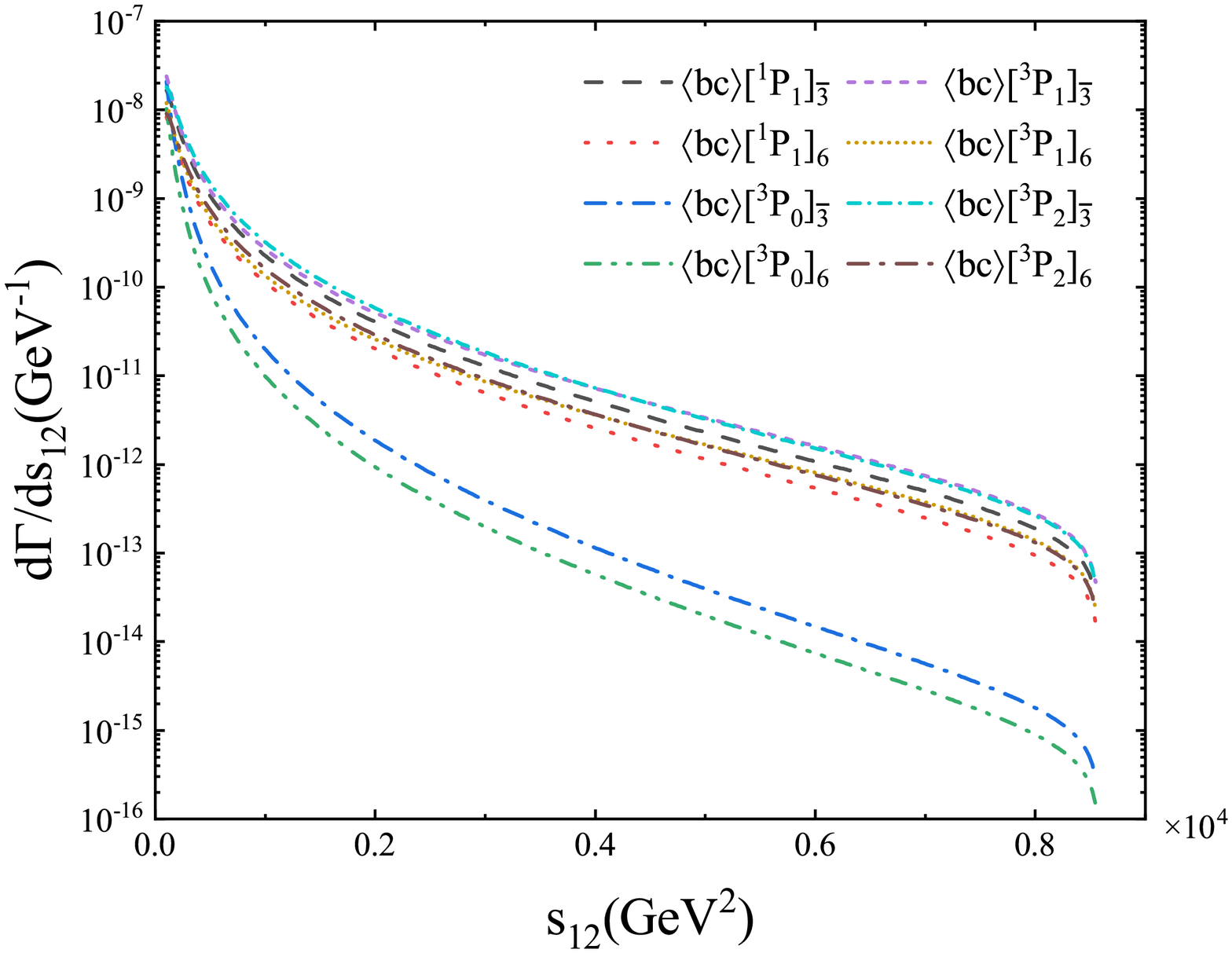}}
  \hspace{0.00in}
  \subfigure[]{
    \includegraphics[width=0.47\textwidth]{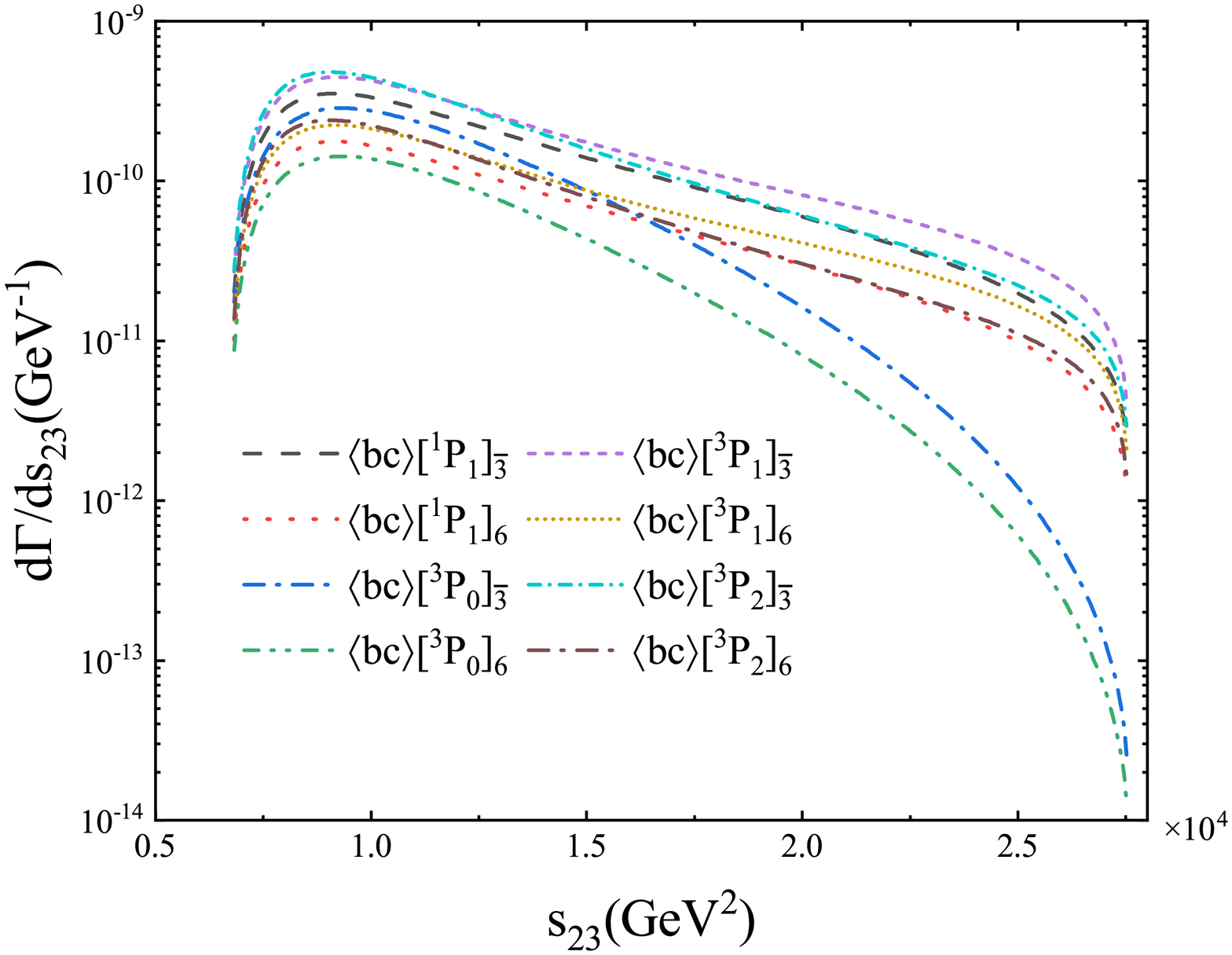}}
  \caption{The differential decay widths $d\Gamma/ds_{12}$ (a) and $d\Gamma/ds_{23}$ (b) for the process $t\rightarrow \Xi_{bc}$[P]+$\bar {Q} + W^{+} $. The dashed black, dotted red, dash-dotted blue, dash-dot-dotted green, short dashed purpe, short dotted yellow, short dash-dotted azury and dash-dotted brown lines represent for the decay widths of P-wave $\Xi_{bc}$ in $[^{1}P_{1}]_{\mathbf{\overline{3}}}$, $ [^{1}P_{1}]_{\mathbf{6}}$, $ [^{3}P_{0}]_{\mathbf{\overline{3}}}$, $ [^{3}P_{0}]_{\mathbf{6}}$, $ [^{3}P_{1}]_{\mathbf{\overline{3}}}$, $ [^{3}P_{1}]_{\mathbf{6}}$, $ [^{3}P_{2}]_{\mathbf{\overline{3}}}$, and $ [^{3}P_{2}]_{\mathbf{6}}$, respectively.}
  \label{s1s2bc}
\end{figure}

\begin{figure}[htb]
  \centering
  \subfigure[]{
    \includegraphics[width=0.47\textwidth]{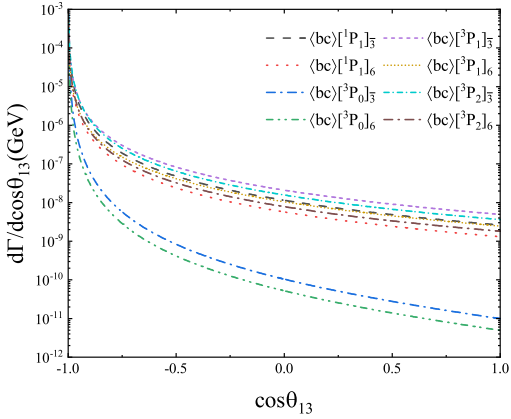}}
  \hspace{0.00in}
  \subfigure[]{
    \includegraphics[width=0.47\textwidth]{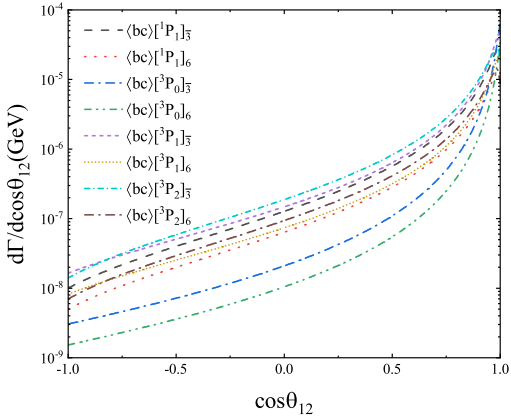}}
  \caption{The differential decay widths $d\Gamma/ \rm{cos}\theta_{13}$ (a) and $d\Gamma/ \rm{cos}\theta_{12}$ (b) for the process $t\rightarrow \Xi_{bc}$[P]+$\bar {Q} + W^{+} $. The dashed black, dotted red, dash-dotted blue, dash-dot-dotted green, short dashed purpe, short dotted yellow, short dash-dotted azury and dash-dotted brown lines represent for the decay widths of P-wave $\Xi_{bc}$ in $[^{1}P_{1}]_{\mathbf{\overline{3}}}$, $ [^{1}P_{1}]_{\mathbf{6}}$, $ [^{3}P_{0}]_{\mathbf{\overline{3}}}$, $ [^{3}P_{0}]_{\mathbf{6}}$, $ [^{3}P_{1}]_{\mathbf{\overline{3}}}$, $ [^{3}P_{1}]_{\mathbf{6}}$, $ [^{3}P_{2}]_{\mathbf{\overline{3}}}$, and $ [^{3}P_{2}]_{\mathbf{6}}$, respectively.}
  \label{cosbc}
\end{figure}

In order to show the characteristics of the excited $\Xi_{bc}$ production through intermediate P-wave diquark $\langle bQ\rangle[n]$ and be helpful for the experiments measurements, we make a clear analysis about the differential distributions in the following, such as $d\Gamma/ds_{12}$, $d\Gamma/ds_{23}$, $d\Gamma/d \rm{cos} \it \theta_{13}$ and $d\Gamma/d \rm{cos} \theta_{12}$, which are shown in Figs.~\ref{s1s2bc} and \ref{cosbc}. Here  the kinematics parameter invariant mass $s_{ij}=(q_{i}+q_{j})^2$, and $\theta_{ij}$ is the angle between outgoing three momenta $\overrightarrow{q_{i}}$ and $\overrightarrow{q_{j}}$ in the rest frame of top quark. In the Figs.~\ref{s1s2bc} and \ref{cosbc}, different color and size of dotted lines are used to represent the differential distributions of eight  intermediate diquark states.

Fig.~\ref{s1s2bc}(a) shows that as the invariant mass $s_{12}$ increases, the differential decay width decreases monotonically.
When $s_{12}$ tends to zero, the $\Xi_{bQ}$ baryon and the heavy antiquark $\overline{Q}$ shall move in much closer directions, resulting in a larger decay width.
In Fig.~\ref{s1s2bc}(b), all the curves are relatively flatter than those in Fig.~\ref{s1s2bc}(a).
In Fig.~\ref{cosbc}, we found that the dependence of the differential decay width $d\Gamma/d \rm{cos}{ \theta_{13}}$ ($d\Gamma/d \rm{cos}{\theta_{12}}$) on $\rm{cos} {\theta_{13}}$ ($\rm{cos} {\theta_{12}}$) also shows a monotonic behavior.
Moreover, the value of $d\Gamma/d \rm{cos}{\theta_{13}}$ is largest when $\rm{cos}{\theta_{13}}=-1$, i.e., $\theta_{13}=180^{\rm{o}}$ and the value of $d\Gamma/d \rm{cos}{\theta_{12}}$ is largest when $\rm{cos}{\theta_{12}}=1$, i.e., $\theta_{12}=0^{\rm{o}}$.

\subsection{The production of $\Xi_{bb}$ baryon}

\begin{table}[htb]
\caption{Decay widths (KeV) and $\rm{Br}_{\it bb}[n]$ for the production of P-wave and S-wave $\Xi_{bb}$ via top quark decays. States represent the spin and color state for the intermediate diquark.}
\vspace{0.5cm}
\begin{tabular}{|c||c||c|}
\hline
~~State~~&~Decay width~&~$\rm{Br}_{\it bb}[n]$~ \\
\hline\hline
$[^1S_0]_{\mathbf{6}}$     & 4.77 & $3.20 \times 10^{-6} $   \\
\hline
$[^3S_1]_{\overline{\mathbf{3}}}$     & 9.37 &  $6.29 \times 10^{-6}$   \\
\hline
$[^1P_1]_{\overline{\mathbf{3}}}$     & 0.125 & $8.39\times10^{-8}$ \\
\hline
$[^3P_0]_{\mathbf{6}}$     & 0.102 & $6.84\times10^{-8}$  \\
\hline
$[^3P_1]_{\mathbf{6}}$     & 0.104 & $6.95\times10^{-8}$ \\
\hline
$[^3P_2]_{\mathbf{6}}$     & 0.035 & $2.33\times10^{-8}$  \\
\hline
S-wave    & 14.14 & $9.49\times10^{-6}$ \\
\hline
P-wave     & 0.365 & $2.45\times10^{-7}$ \\
\hline
Total    & 14.505 & $9.73 \times10^{-6}$ \\
\hline
\end{tabular}
\label{bbwidth}
\end{table}

Similar to the production of $\Xi_{bc}$, the decay widths for all considered spin and color configurations through the process $t\rightarrow \Xi_{bb}+ \bar {b}+W^{+}$ are listed in Table~\ref{bbwidth}, which show that
\begin{itemize}
  \item The total contributions from the P-wave $\Xi_{bb}$ is about 2\% of that from P-wave $\Xi_{bc}$. Because $m_b$ is about three times larger than $m_c$, the phase space of $\Gamma_{t \rightarrow \Xi_{bb}[P]}$ is suppressed compared with that of $\Gamma_{t \rightarrow \Xi_{bc}[P]}$.

  \item All of the four spin and color states for the production of excited $\Xi_{bb}$ have been considered in our numerical calculation and the ratio among them is $[^{1} P _1]_{\bar {\mathbf{3}}}:[^{3} P _1]_{\mathbf{6}}:[^{3} P _0]_{\mathbf{6}}:[^{3} P _2]_{\mathbf{6}}=1:0.83:0.81:0.28$.

  \item For the production of excited $\Xi_{bb}$, the contribution from the spin and color state $[^{1} P _1]_{\bar {\mathbf{3}}}$ domains. The total contribution from intermediate P-wave diquark state is about 2.58\% of that from S-wave.

  \item At the LHC, about $9.73 \times 10^{2-4}$ $\Xi_{bb}$ events/yr will be produced with the sum of S-wave and P-wave states via top-quark decays. The total proportion coming from the P-wave state is about $2.45 \times 10^{1-3}$ events/yr. The events per year for the $[^{1} P _1]_{\bar {\mathbf{3}}}$, $[^{3} P _0]_{\mathbf{6}}$, $[^{3} P _1]_{\mathbf{6}}$ and $[^{3} P _2]_{\mathbf{6}}$ are about $8.39 \times 10^{0-2}$, $6.84 \times 10^{0-2}$, $6.95 \times 10^{0-2}$, and $2.33 \times 10^{0-2}$, accordingly.

\end{itemize}

\begin{figure}[htb]
  \centering
  \subfigure[]{
    \includegraphics[width=0.47\textwidth]{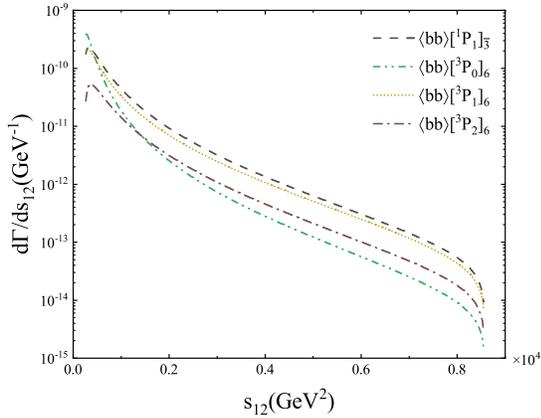}}
  \hspace{0.00in}
  \subfigure[]{
    \includegraphics[width=0.47\textwidth]{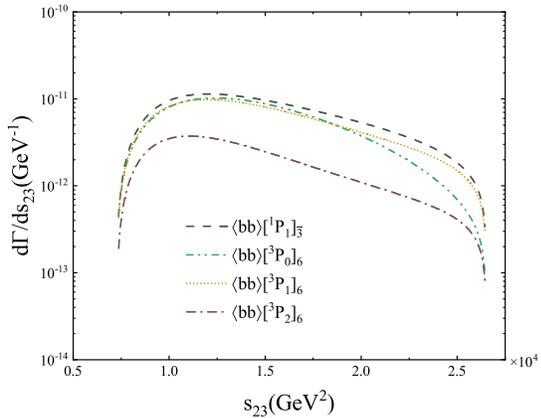}}
 \caption{The differential decay widths $d\Gamma/ds_{12}$ (a) and $d\Gamma/ds_{23}$ (b) for the process $t\rightarrow \Xi_{bb}$[P]+$\overline {Q} + W^{+} $. The dashed black, dash-dot-dotted green, short dotted yellow and dash-dotted brown lines represent for the decay widths of P-wave $\Xi_{bb}$ in $[^{1}P_{1}]_{\mathbf{\overline{3}}}$, $ [^{3}P_{0}]_{\mathbf{6}}$, $ [^{3}P_{1}]_{\mathbf{6}}$, and $ [^{3}P_{2}]_{\mathbf{6}}$, respectively.}
  \label{s1s2bb}
\end{figure}

\begin{figure}[htb]
  \centering
  \subfigure[]{
    \includegraphics[width=0.47\textwidth]{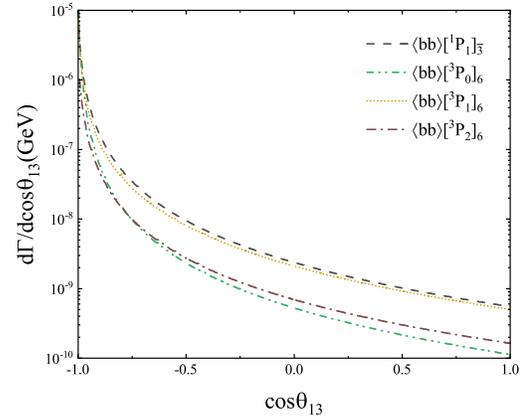}}
  \hspace{0.00in}
  \subfigure[]{
    \includegraphics[width=0.47\textwidth]{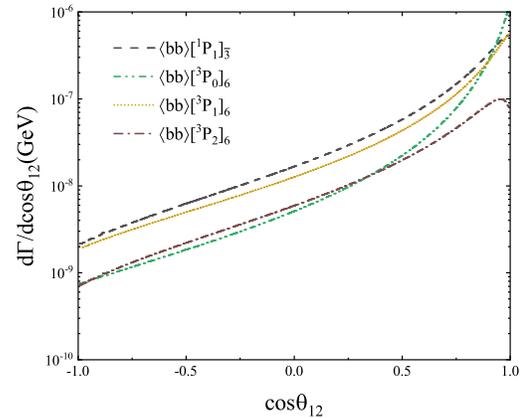}}
     \caption{The differential decay widths $d\Gamma/ \rm{cos}\theta_{13}$ (a) and $d\Gamma/ \rm{cos}\theta_{12}$ (b) for the process $t\rightarrow \Xi_{bb}$[P]+$\overline {Q} + W^{+} $. The dashed black, dash-dot-dotted green, short dotted yellow and dash-dotted brown lines represent for the decay widths of P-wave $\Xi_{bb}$ in $[^{1}P_{1}]_{\mathbf{\overline{3}}}$, $ [^{3}P_{0}]_{\mathbf{6}}$, $ [^{3}P_{1}]_{\mathbf{6}}$, and $ [^{3}P_{2}]_{\mathbf{6}}$, respectively.}
       \label{cosbb}
\end{figure}

The differential distributions, such as $d\Gamma/ds_{12}$, $d\Gamma/ds_{23}$, $d\Gamma/d \rm{cos} \theta_{13}$, and $d\Gamma/d \rm{cos} \theta_{12}$, for the production of $\Xi_{bb}$ through the process $t(p_1) \rightarrow \Xi_{bb} (q_1)+ \bar {b} (q_2) + W^{+} (q_3)$ are shown in Figs.~\ref{s1s2bb} and \ref{cosbb}. The kinematics parameters $s_{ij}$ and $\theta_{ij}$ are defined the same as that in Sec. \ref{secbc}. From Figs.~\ref{s1s2bb} and \ref{cosbb}, more characteristics can be found to be basicaly consistent with that for the production of $\Xi_{bc}$.

\subsection{Theoretical uncertainty}

In this subsection, the theoretical uncertainty for the production of $\Xi_{bQ}$ shall be discussed.
There are three main sources of theoretical uncertainty, namely mass parameters, transition probability and the renormalization scale.

\begin{table}[htb]
\centering
\caption{The theoretical uncertainty for the production of $\Xi_{bc}$ (in unit. $10^{-7}$ GeV) caused by charm quark mass. Here we vary $m_c$ over the range [1.5, 2.1] GeV while the masses of other two heavy quarks are taken as their central values, namely, $m_b=5.1~\rm{GeV}$ and $m_t=173.0~\rm{GeV}$.} \vspace{0.5cm}
\begin{tabular}{|c|c|c|c|c|c|} 
\hline
$m_c$(GeV)                                           & 1.50~  & 1.65~  & 1.80~  & 1.95~  & 2.10~   \\ 
\hline\hline
$\Gamma_{\Xi_{bc}}([^1P_1]_{\overline{\mathbf{3}}})$ & 65.96~ & 41.23~ & 26.90~ & 18.20~ & 12.69~  \\ 
\hline
$\Gamma_{\Xi_{bc}}([^1P_1]_{\mathbf{6}})$            & 32.98~ & 20.61~ & 13.45~ & 9.10~  & 6.35~   \\ 
\hline
$\Gamma_{\Xi_{bc}}([^3P_0]_{\overline{\mathbf{3}}})$ & 38.14~ & 25.85~ & 18.18~ & 13.18~ & 9.81~   \\ 
\hline
$\Gamma_{\Xi_{bc}}([^3P_0]_{\mathbf{6}})$            & 19.07~ & 12.93~ & 9.09~  & 6.59~  & 4.90~   \\ 
\hline
$\Gamma_{\Xi_{bc}}([^3P_1]_{\overline{\mathbf{3}}})$ & 80.98~ & 51.89~ & 34.63~ & 23.92~ & 17.01~  \\ 
\hline
$\Gamma_{\Xi_{bc}}([^3P_1]_{\mathbf{6}})$            & 40.49~ & 25.94~ & 17.32~ & 11.96~ & 8.51~   \\ 
\hline
$\Gamma_{\Xi_{bc}}([^3P_2]_{\overline{\mathbf{3}}})$ & 89.35~ & 53.46~ & 33.41~ & 21.67~ & 14.50~  \\ 
\hline
$\Gamma_{\Xi_{bc}}([^3P_2]_{\mathbf{6}})$            & 44.68  & 26.73  & 16.71  & 10.83  & 7.25   \\ 
\hline
\end{tabular}
\label{mcun}
\end{table}

\begin{table}[htb]
\centering
\caption{The theoretical uncertainty for the production of $\Xi_{bQ}$ (in unit. $10^{-7}$ GeV) caused by bottom quark mass. Here we vary $m_b$ over the range [4.7, 5.5] GeV while the masses of other two heavy quarks are taken as their central values, namely, $m_c=1.8~\rm{GeV}$ and $m_t=173.0~\rm{GeV}$.} \vspace{0.5cm}
\begin{tabular}{|c|c|c|c|c|c|} 
\hline
$m_b$(GeV)                                           & 4.7    & 4.9    & 5.1    & 5.3    & 5.5     \\ 
\hline\hline
$\Gamma_{\Xi_{bc}}([^1P_1]_{\overline{\mathbf{3}}})$ & 27.67~ & 27.27~ & 26.90~ & 26.55~ & 26.23~  \\ 
\hline
$\Gamma_{\Xi_{bc}}([^1P_1]_{\mathbf{6}})$            & 13.84~ & 13.64~ & 13.45~ & 13.28~ & 13.12~  \\ 
\hline
$\Gamma_{\Xi_{bc}}([^3P_0]_{\overline{\mathbf{3}}})$ & 19.87~ & 18.98~ & 18.18~ & 17.45~ & 16.77~  \\ 
\hline
$\Gamma_{\Xi_{bc}}([^3P_0]_{\mathbf{6}})$            & 9.93~  & 9.49~  & 9.09~  & 8.72~  & 8.39~   \\ 
\hline
$\Gamma_{\Xi_{bc}}([^3P_1]_{\overline{\mathbf{3}}})$ & 36.30~ & 35.44~ & 34.63~ & 33.89~ & 33.20~  \\ 
\hline
$\Gamma_{\Xi_{bc}}([^3P_1]_{\mathbf{6}})$            & 18.15~ & 17.72~ & 17.32~ & 16.94~ & 16.60~  \\ 
\hline
$\Gamma_{\Xi_{bc}}([^3P_2]_{\overline{\mathbf{3}}})$ & 33.01~ & 33.22~ & 33.41~ & 33.59~ & 33.75~  \\ 
\hline
$\Gamma_{\Xi_{bc}}([^3P_2]_{\mathbf{6}})$            & 16.51~ & 16.61~ & 16.71~ & 16.79~ & 16.87   \\ 
\hline\hline
$\Gamma_{\Xi_{bb}}([^1P_1]_{\overline{\mathbf{3}}})$ & 1.95~  & 1.55~  & 1.25~  & 1.01~  & 0.82~   \\ 
\hline
$\Gamma_{\Xi_{bb}}([^3P_0]_{\mathbf{6}})$            & 1.56~  & 1.26~  & 1.02~  & 0.83~  & 0.68~   \\ 
\hline
$\Gamma_{\Xi_{bb}}([^3P_1]_{\mathbf{6}})$            & 1.61~  & 1.29~  & 1.04~  & 0.84~  & 0.69~   \\ 
\hline
$\Gamma_{\Xi_{bb}}([^3P_2]_{\mathbf{6}})$            & 0.55~  & 0.43~  & 0.35~  & 0.28~  & 0.23~   \\
\hline
\end{tabular}
\label{mbun}
\end{table}

\begin{table}[htb]
\centering
\caption{The theoretical uncertainty for the production of $\Xi_{bQ}$ (in unit. $10^{-7}$ GeV) caused by top quark mass. Here we vary $m_t$ over the range [172.6, 173.4] GeV while the masses of other two heavy quarks are taken as their central values, namely, $m_b=5.1~\rm{GeV}$ and $m_c=1.8~\rm{GeV}$.} \vspace{0.5cm}
\begin{tabular}{|c|c|c|c|c|c|} 
\hline
$m_t$(GeV)                                           & 172.6~ & 172.8~ & 173.0~ & 173.2~ & 173.4~  \\ 
\hline\hline
$\Gamma_{\Xi_{bc}}([^1P_1]_{\overline{\mathbf{3}}})$ & 26.66~ & 26.78~ & 26.90~ & 27.02~ & 27.14~  \\ 
\hline
$\Gamma_{\Xi_{bc}}([^1P_1]_{\mathbf{6}})$            & 13.33~ & 13.39~ & 13.45~ & 13.51~ & 13.57~  \\ 
\hline
$\Gamma_{\Xi_{bc}}([^3P_0]_{\overline{\mathbf{3}}})$ & 18.03~ & 18.10~ & 18.18~ & 18.26~ & 18.33~  \\ 
\hline
$\Gamma_{\Xi_{bc}}([^3P_0]_{\mathbf{6}})$            & 9.01~  & 9.05~  & 9.09~  & 9.13~  & 9.17~   \\ 
\hline
$\Gamma_{\Xi_{bc}}([^3P_1]_{\overline{\mathbf{3}}})$ & 34.33~ & 34.48~ & 34.63~ & 34.79~ & 34.94~  \\ 
\hline
$\Gamma_{\Xi_{bc}}([^3P_1]_{\mathbf{6}})$            & 17.16~ & 17.24~ & 17.32~ & 17.39~ & 17.47~  \\ 
\hline
$\Gamma_{\Xi_{bc}}([^3P_2]_{\overline{\mathbf{3}}})$ & 33.11~ & 33.26~ & 33.41~ & 33.56~ & 33.72~  \\ 
\hline
$\Gamma_{\Xi_{bc}}([^3P_2]_{\mathbf{6}})$            & 16.55  & 16.63  & 16.71  & 16.78  & 16.86   \\ 
\hline\hline
$\Gamma_{\Xi_{bb}}([^1P_1]_{\overline{\mathbf{3}}})$ & 1.24~  & 1.24~  & 1.25~  & 1.26~  & 1.26~   \\ 
\hline
$\Gamma_{\Xi_{bb}}([^3P_0]_{\mathbf{6}})$            & 1.01~  & 1.01~  & 1.02~  & 1.02~  & 1.03~   \\ 
\hline
$\Gamma_{\Xi_{bb}}([^3P_1]_{\mathbf{6}})$            & 1.03~  & 1.03~  & 1.04~  & 1.04~  & 1.05~   \\ 
\hline
$\Gamma_{\Xi_{bb}}([^3P_2]_{\mathbf{6}})$            & 0.34~  & 0.34~  & 0.35~  & 0.35~  & 0.35~   \\
\hline
\end{tabular}
\label{mtun}
\end{table}

The theoretical uncertainty caused by the mass of heavy quark ($m_c$, $m_b$ and $m_t$) and $W^+$ boson would be given priority consideration. For the production of $\Xi_{bc}$, heavy quark masses $m_c$, $m_b$ and $m_t$ are all the error sources of theoretical uncertainties, which are shown in Tables~\ref{mcun},~\ref{mbun} and~\ref{mtun}, respectively. However, as for the production of $\Xi_{bb}$, there are only two main sources of heavy quark mass uncertainty, i.e. $m_b$ and $m_t$, and they are also shown in Tables~\ref{mbun} and~\ref{mtun}, respectively. To estimate the uncertainties, we adopt $m_c=1.8\pm0.3~\rm{GeV}$, $m_b=5.1\pm0.4~\rm{GeV}$, and $m_t=173.0\pm0.4~\rm{GeV}$. From Tables~\ref{mcun}-\ref{mtun}, one can find that:
\begin{itemize}
  \item The decay width for the production of excited $\Xi_{bc}$ baryon is more sensitive to $m_c$ than $m_b$ and $m_t$. Because the suppression of phase space, $\Gamma_{\Xi_{bc}[n]}$ decreases with the increment of $m_c$. Due to the influence of the projector in Eq.~(\ref{piall}) and the polarization tensors sum formula in Eqs.~(\ref{psum3p0}-\ref{psum3p2}), the decay width of excited $\Xi_{bc}$ in $[^3 P_2]_{\mathbf{\overline 3}/ \mathbf{6}}$ state increases with the increment of $m_b$, while others decrease with the increment of $m_b$.
  \item For the production of excited $\Xi_{bb}$, the decay width is more sensitive to $m_b$ than $m_t$, and the uncertainty effect from $m_t$ is small enough to be ignored. The decay width decreases with the increment of $m_b$ for the suppression of phase space.
\end{itemize}
Based on the recent difference of the W boson mass with a significance of 7.0$\sigma$ between the CDF-II and the standard model~\cite{CDF:2022hxs}, we also calculate its mass uncertainty, and finally find that it has a negligible impact on the final results. 
Therefore, the theoretical uncertainty for the production of $\Xi_{bQ}$ caused by $W^+$ boson mass would not be shown in this paper.

Then the theoretical uncertainty caused by the transition probability $h_{\mathbf{6}}$ shall be discussed. Based on those two different points of view mentioned in Sec.~\ref{hadronization} about the contributions of diquark in color $\mathbf{\bar 3}$ and $\mathbf{6}$ state, the total produced $\Xi_{bQ}$ events would be roughly estimate at the LHC in one operetion year with $h_{\mathbf{6}}=0$, or $h_{\mathbf{6}}=v^{2} h_{\mathbf{\overline{3}}}$ with $v^2=(0.1-0.3)$. And the corresponding results, along with those obtained with $h_{\mathbf{6}}=h_{\mathbf{\overline{3}}}$ for comparison, are presented in Table~\ref{utp}. Nevertheless, the uncertainty effect from the transition probability can be easily estimated when a relatively reliable value of $h_\mathbf{6}$ appears, because it is proportional to the decay width for the production of doubly heavy baryon $\Xi_{bQ}$ within the NRQCD framework.

\begin{table}[htb]
\centering
\caption{The theoretical uncertainty for the total produced $\Xi_{bQ}$ events caused by the transition probability $h_{\mathbf{6}}$ at LHC with a high Luminosity $\mathcal{L}=10^{34-36} \rm ~cm^{-2}s^{-1}$ in one operation year. Here we take $h_{\mathbf{6}}=0$, $h_{\mathbf{6}}=v^{2} h_{\mathbf{\overline{3}}}$, or $h_{\mathbf{6}}=h_{\mathbf{\overline{3}}}$, respectively, while the masses of heavy quarks are taken as their central values, namely, $m_b=5.1~\rm{GeV}$,  $m_c=1.8~\rm{GeV}$ and $m_t=173.0~\rm{GeV}$.} \vspace{0.5cm}
\begin{tabular}{|c|c|c|c|} 
\hline
Events  & $h_{\mathbf{6}}=0$ & $h_{\mathbf{6}}=v^{2} h_{\mathbf{\overline{3}}} $~& $h_{\mathbf{6}}=h_{\mathbf{\overline{3}}} $~  \\ 
\hline\hline
$N_{\Xi_{bc}}~(\times 10^{4-6})$ & 1.58~ & 1.66-1.81~ & $2.36$~  \\ 
\hline
$N_{\Xi_{bb}}~(\times 10^{2-4})$ & 6.37~ & 6.71-7.38~ & $9.73$~  \\ 
\hline
\end{tabular}
\label{utp}
\end{table}

Finally, uncertainty induced by the choices of renormalization scale $\mu_r$ is presented in Table~\ref{urun}, in which three different renormalization scales $\mu_r=2m_c$, $\rm{M}_{bc}$, and $2m_b$ are used both for the production of excited $\Xi_{bc}$ and $\Xi_{bb}$.
Obviously, the selection of renormalization scale $\mu_r$ can cause great uncertainty. However, such a scale ambiguity could be suppressed by a higher-order perturbative calculation or principle of maximum conformal scale-setting method (PMC) \cite{Brodsky:2011ta, Brodsky:2012rj, Mojaza:2012mf, Brodsky:2013vpa,Wu:2013ei,Wu:2014iba}.
Fortunately, different renormalization scales only affect the strong coupling constant, which is proportional to the decay width. So its uncertainty effect can be easily estimated.

\begin{table}[htb]
\centering
\caption{The theoretical uncertainty for the production of $\Xi_{bQ}$ (in unit. $10^{-7}$ GeV) caused by the renormalization scale $\mu_r$. Here we take $\mu_r=2m_c$, $\rm{M}_{bc}$, or $2m_b$, respectively, while the masses of heavy quarks are taken as their central values, namely, $m_b=5.1~\rm{GeV}$,  $m_c=1.8~\rm{GeV}$ and $m_t=173.0~\rm{GeV}$.} \vspace{0.5cm}
\begin{tabular}{|c|c|c|c|} 
\hline
$\mu_r$(GeV)                                           & $2m_c$    & $\rm{M}_{\it bc}$    & $2m_b$     \\ 
\hline\hline
$\Gamma_{\Xi_{bc}}([^1P_1]_{\overline{\mathbf{3}}})$ & 26.90~ & 18.09~ & 14.92~  \\ 
\hline
$\Gamma_{\Xi_{bc}}([^1P_1]_{\mathbf{6}})$            & 13.45  & 9.05   & 7.46    \\ 
\hline
$\Gamma_{\Xi_{bc}}([^3P_0]_{\overline{\mathbf{3}}})$ & 18.18~ & 12.23~ & 10.08~  \\ 
\hline
$\Gamma_{\Xi_{bc}}([^3P_0]_{\mathbf{6}})$            & 9.09   & 6.11   & 5.04    \\ 
\hline
$\Gamma_{\Xi_{bc}}([^3P_1]_{\overline{\mathbf{3}}})$ & 34.63~ & 23.29~ & 19.21~  \\ 
\hline
$\Gamma_{\Xi_{bc}}([^3P_1]_{\mathbf{6}})$            & 17.32  & 11.65  & 9.61    \\ 
\hline
$\Gamma_{\Xi_{bc}}([^3P_2]_{\overline{\mathbf{3}}})$ & 33.41~ & 22.47~ & 18.53~  \\ 
\hline
$\Gamma_{\Xi_{bc}}([^3P_2]_{\mathbf{6}})$            & 16.71  & 11.24  & 9.27    \\ 
\hline\hline
$\Gamma_{\Xi_{bb}}([^1P_1]_{\overline{\mathbf{3}}})$ & 2.25~  & 1.51~  & 1.25~   \\ 
\hline
$\Gamma_{\Xi_{bb}}([^3P_0]_{\mathbf{6}})$            & 1.84~  & 1.24~  & 1.02~   \\ 
\hline
$\Gamma_{\Xi_{bb}}([^3P_1]_{\mathbf{6}})$            & 1.87~  & 1.26~  & 1.04~   \\ 
\hline
$\Gamma_{\Xi_{bb}}([^3P_2]_{\mathbf{6}})$            & 0.63~  & 0.42~  & 0.35~   \\
\hline
\end{tabular}
\label{urun}
\end{table}

\section{Summary}
In this paper, the production of excited doubly heavy baryons $\Xi_{bQ}$ are analyzed through the process $ t\rightarrow\Xi_{bQ}+\overline{Q}+W^+$ ($Q$=$b$ or $c$) within the framework of NRQCD. In the numerical calculation, all the possible intermediate P-wave diquark states with spin and color quantum number, i.e., $\langle bc\rangle[^1 P_1]_{\overline{\mathbf{3}}/\mathbf{6}}$, $\langle bc\rangle[^3 P_J]_{\overline{\mathbf{3}}/\mathbf{6}}$, $\langle bb\rangle[^1 P_1]_{\overline{\mathbf{3}}}$ and $\langle bb\rangle[^3 P_J]_{\mathbf{6}}$ have been taken into consideration. By summing up all the contributions from these diquark states, the decay width for the P-wave $\Xi_{bc}$ and $\Xi_{bb}$ are 16.96~KeV and 0.365~KeV, respectively. We observe that the contribution from P-wave states is 5.05\%(2.58\%) of that from S-wave states for the production of $\Xi_{bc}$($\Xi_{bb}$).
The contribution of P-wave can be regarded as the higher-order contribution of S-wave. The differential distributions, such as $d\Gamma/ds_{12}$, $d\Gamma/ds_{23}$, $d\Gamma/d \rm{cos} \theta_{13}$, and $d\Gamma/d \rm{cos} \theta_{12}$, are presented to show the characteristics of this process for the excited $\Xi_{bc}$ and $\Xi_{bb}$ production and be helpful for the experiments measurements. From these distributions, it can be found that when the outgoing $\Xi_{bQ}$ and antiquark $\overline{Q}$ move in the same direction, and both of them move back to back with $W^{+}$ boson, the differential decay width can achieve its largest value.

The decay width coming from the intermediate P-wave diquark states can be obtained in the following,
\begin{eqnarray}
&&\Gamma_{t\rightarrow \Xi_{bc}[P]+\bar{c}+W^+}=16.96^{+25.64}_{-12.66} ~\rm{KeV},\nonumber \\
&&\Gamma_{t\rightarrow \Xi_{bb}[P]+\bar{b}+W^+}=0.365^{+0.668}_{-0.124} ~\rm{KeV},\nonumber
\end{eqnarray}
where the uncertainties come from the heavy quark masses ($m_b,~m_c$ and $m_t$), and the renormalization scale $\mu_r$.
The decay width for the production of $\Xi_{bc}$ ($\Xi_{bb}$) is more sensitive to $m_c$ ($m_b$) than other sources of the theoretical uncertainty, which is mainly caused by the change of phase space. 
Because of the running behavior of strong coupling constant $\alpha_s(\mu_r)$, the selection of the renormalization scale $\mu_r$ has a big difference in the decay width through this process. Fortunately, the decay width at the leading order is proportional to the strong coupling constant $\alpha_s$ and the uncertainty effect induced by the renormalization scale $\mu_r$ can be estimated easily.

If all these excited P-wave doubly heavy baryons decay into the ground state completely, we can obtain the total decay width
\begin{eqnarray}
&&\Gamma_{t\rightarrow \Xi_{bc}+\bar{c}+W^+}=352.66^{+306.14}_{-237.36} ~\rm{KeV},\nonumber \\
&&\Gamma_{t\rightarrow \Xi_{bb}+\bar{b}+W^+}=14.50^{+20.33}_{-3.42} ~\rm{KeV}.\nonumber
\end{eqnarray}
In this case of a combination of S- and P-wave contributions, there will be about $2.36\times 10^{4-6}$ $\Xi_{bc}$ events and $9.73\times 10^{2-4}$ $\Xi_{bb}$ events produced in one operation year at the LHC with a high luminosity $\mathcal{L}=10^{34-36} \rm  ~cm^{-2}s^{-1}$. Therefore, there are still sizable events of $\Xi_{bc}$ and $\Xi_{bb}$ baryons produced at the LHC in one operation year. With so many events, the experimental signals will possibly be found at the LHC in the future. And at the future $e^{+} e^{-}$ colliders with clean background, high luminosity and high collision energy, a huge number of top quark will be produced. With the help of Eq.~(\ref{Nt}) and our results, the events for the produced $\Xi_{bc}$ and $\Xi_{bb}$ baryons can be estimated at the future $e^{+} e^{-}$ colliders.

It is worth mentioning that the isospin-breaking effect is ignored in this paper. That is to say $\Xi_{bc}$ denotes $\Xi^{+}_{bcu}$, $\Xi^{0}_{bcd}$ and $\Omega^{0}_{bcs}$ and $\Xi_{bb}$ denotes $\Xi^{0}_{bbu}$, $\Xi^{-}_{bbd}$ and $\Omega^{-}_{bbs}$. Because the intermediate diquark state $\langle b Q \rangle$ needs to grab a light $u$, $d$, or $s$ quark in the vacuum to form the doubly heavy baryon $\Xi_{b Q}$ accordingly. According to the ratio for the production of $\Xi_{b Qu}$, $\Xi_{b Qd}$ and $\Omega_{b Qs}$ is 1:1:0.3 \cite{Sjostrand:2006za}, there will be about 43\% $\Xi_{bc}$ events become $\Xi_{bc}^{+}$, 43\% to $\Xi_{bc}^{0}$, and 14\% to $\Omega_{bc}^{0}$, and the same percentage for $\Xi_{bc}$ events become $\Xi_{bb}^{0}$, $\Xi_{bb}^{-}$, and $\Omega_{bb}^{-}$. 

\hspace{2cm}

{\bf Acknowledgements}: This work was partially supported by the Central Government Guidance Funds for Local Scientific and Technological Development, China (No. Guike ZY22096024). This work was also supported by the National Natural Science Foundation of China (Grants No. 12005045, No. 12047506 and No. 12005028). This research was also supported by the Project funded by China Postdoctoral Science Foundation (No. 2021M690234) and Natural Science Foundation of Guangxi (No. 2020GXNSFBA159003).

\bibliographystyle{physrev}
\bibliography{doubleheavy.bib}

\end{document}